\documentclass[11pt,a4paper,twoside]{article}
\usepackage[T1]{fontenc}
\usepackage[utf8x]{inputenc}
\usepackage[italian,english]{babel}
\usepackage{url}
\usepackage{ucs}
\usepackage{eurosym}
\usepackage{fancybox}
\usepackage{fancyhdr}
\usepackage{booktabs}
\usepackage{longtable}
\usepackage{mdframed}
\usepackage{graphics}
\usepackage{setspace}
\usepackage{xspace}
\usepackage{graphicx}
\usepackage{tabularx}
\usepackage{multirow}
\usepackage[table]{xcolor}
\usepackage{colortbl}
\usepackage{siunitx}
\usepackage{lettrine}
\usepackage{enumitem}
\usepackage{ulem}
\usepackage{microtype}
\usepackage{alltt}
\usepackage{wrapfig}
\usepackage[format=hang,textformat=period,labelfont={bf}]{caption}
\usepackage{listings}
\usepackage{mathpazo}
\usepackage{charter}
\usepackage{trajan}
\usepackage{courier}
\usepackage{soul}
\usepackage[english]{varioref}

\usepackage[breaklinks=true]{hyperref}
\hypersetup{
    colorlinks  = true,
    linktoc     = all,
    unicode     = true,
    citecolor   = royalpurple,
    filecolor   = blue,
    linkcolor   = blue,
    urlcolor    = red,
}

\renewcommand{\emph}[1]{\textit{#1}}

\begin{document}

\author{Michele Finelli}

\title{How many years can a tiny unbalanced parenthesis go unnoticed on
a widely accessed Internet document, older than the \textit{World Wide
Web} itself ?}

\date{\today}

\maketitle

\begin{abstract}
    26
\end{abstract}

\section{Introduction} 
\label{sec:Introduction}

Reading the RFC Index\footnote{As everybody does on their Christmas
holidays, at least in civilized lands.} we noticed a syntax error: a
tiny unbalanced parenthesis on one of the first paragraphs, as reported
on table \vref{table:comparison}. For example, the file available at
\url{https://www.ietf.org/rfc/rfc-index} still has this issue, as of
today.

\begin{table}
    \begin{center}
        \begin{tabularx}{\linewidth}{lX}
            \toprule
            \textbf{Wrong} & 
            Obsoletes xxxx refers to other RFCs that this one replaces;
    Obsoleted by xxxx refers to RFCs that have replaced this one.
    Updates xxxx refers to other RFCs that this one merely updates 
    \textit{but does not replace)}; Updated by xxxx refers to RFCs that
    have updated (but not replaced) this one.  Generally, only
    immediately succeeding and/or preceding RFCs are indicated, not the
    entire history of each related earlier or later RFC in a related
    series. \\
            \midrule
            \textbf{Right} &
            Obsoletes xxxx refers to other RFCs that this one replaces;
    Obsoleted by xxxx refers to RFCs that have replaced this one.
    Updates xxxx refers to other RFCs that this one merely updates
    \textbf{(but does not replace)}; Updated by xxxx refers to RFCs that
    have updated (but not replaced) this one.  Generally, only
    immediately succeeding and/or preceding RFCs are indicated, not the
    entire history of each related earlier or later RFC in a related
    series. \\
            \bottomrule
        \end{tabularx}
        \caption{The sentence is \textit{italic} in the first block and
        \textbf{bold} in the second}
        \label{table:comparison}
    \end{center}
\end{table}

The error is clearly a minor one, but we realized it was present also in
older releases of the same document, for example looking at the
\textit{Wayback machine} of \url{https://web.archive.org} we checked
that the error is already present on the first occurrence, dated 15 June
2007. Also, on the same IETF site, there is a PDF
file\footnote{\url{https://www.ietf.org/rfc/rfc-index.txt.pdf}} dating
back to 2012-04-06 10:17 that shows the same issue --- the PDF, by the
way, reports ``CREATED ON: 04/04/2012.'' so it is probably really
sitting on that web folder since April 2012. 

The RFCs are a fundamental part of the Internet documents, the oldest of
all: RFC 1, titled ``Host Software'', was written by Steve Crocker and
published on 7 April 1969.  We may assume that, since there was a list
of RFCs, there was also a document with the list of the citations to the
published RFCs and that a document like the \texttt{rfc-index.txt} file
(maybe by another name) surely existed since the '70s.  Because of the
relevance of the RFCs we may also assume that the RFC index has been
downloaded \textit{a lot} of times --- but probably not \textit{read} in
detail many times, as the present investigation suggests.

This brings to the question that gives the title to the paper: ``How
many years can a tiny unbalanced parenthesis go unnoticed on a widely
accessed Internet document, older than \textit{WWW} itself ?'' (probably
\textit{much older} than the WWW.)


\section{The quest for the RFC indexes}
\label{sec:The quest for the RFC indexes}

We began our quest for \textit{ancient} RFC indexes: we wanted to find
the precise date where the error was introduced, or the best estimate we
could determine.

The main issue is that, while a web search for a \texttt{rfc-index.txt}
file returns millions of hits, what we really need are some of the very
few \textit{archived} versions, not one of the many recent ones.

An help came by noticing that the \texttt{rfc-index.txt} has some
``brother and sisters'' documents, as stated in RFC2648: the \textit{For
Your Information} \texttt{fyi-index.txt}, the \textit{Standards}
\texttt{std-index.txt} and the \textit{Best Common Practices}
\texttt{bcp-index.txt}. Moreover, and more important for our
investigation, current releases of those files show the same mistake,
with the exception of the \texttt{std-index.txt} files, because they do
not have that paragraph. Besides, these three indexes (RFC, FYI and BCP)
show a very similar structure, and in particular the sentence
``Obsoletes xxxx (\dots) in a related series.'' is the most conserved
region of the header --- clearly they differ in the second part of the
file, where the proper citations list begins. So, it seems reasonable to
infer that they are generated by some software\footnote{We have
not been able to track the source.} that makes the same mistake whenever
it builds the indexes, and that we could use dates and references of
either index to pin down the time of change.

To cut a long story short, we believe that there is good evidence that
the change happened between the 14 and the 15 of July 1994: the key
finding was the FTP server of the Clausthal University of Technology
(see the Resources section \vref{sec:Resources} for the URLs), which
contains some indexes that seems to mark exactly the threshold between
the latest occurrence of a correct text and the earliest occurrence of
the mistake.


\section{Analysis} 
\label{sec:Analysis}

We downloaded the files at the following URLs and analyzed their content
and metadata (for readability, the path is relative to
\url{ftp://ftp.tu-clausthal.de/pub/docs/rfc/}):

\begin{description}

    \item[rfc-index.txt] \texttt{/other\_indexes/rfc-index.txt}

    \item[fyi-index.txt] \texttt{/other\_indexes/fyi-index.txt}

    \item[std-index.txt] \texttt{/other\_indexes/std-index.txt} which is
        in fact a symbolic link for the file
        \texttt{/standards/std-index.txt}

\end{description}

Table \vref{table:indexesmetadata} shows the file size and the ISO 8601
date format of the three files. The reader may notice a strange thing,
namely that the \texttt{fyi-index.txt} and the \texttt{std-index.txt}
have exactly the same size, and this is very suspicious since their
content should be quite different. Looking at the content the mystery
unveils: all the indexes are in fact RFC indexes, despite the file name
says otherwise.

\begin{table}
    \begin{center}
        \begin{tabularx}{\linewidth}{llX}
            \toprule
            \textbf{Filename} & \textbf{Bytes} & \textbf{Date - ISO 8601 format} \\
            \midrule
            fyi-index.txt & 186821 & 1994-07-15 20:58:34.000000000 +0200 \\
            rfc-index.txt & 222733 & 1994-07-15 21:32:32.000000000 +0200 \\
            std-index.txt & 186821 & 1994-07-15 21:38:13.000000000 +0200 \\
            \bottomrule
        \end{tabularx}
        \caption{Metadata of index files downloaded from TU Clausthal's FTP
        server}
        \label{table:indexesmetadata}
    \end{center}
\end{table}

An MD5 comparison of \texttt{fyi-index.txt} and \texttt{std-index.txt}
shows that the files are indeed different. An inspection of the files
shows that they are the same until line 271: after that the
\texttt{fyi-index.txt} has a sequence of \texttt{\^{}@} characters, as
if it was corrupted (it is worth to mention that also the
\texttt{rfc-index.txt} file is similarly corrupted, beginning from line
232).

There seems to be two different RFC index files: the first is properly
named \texttt{rfc-index.txt} while the second is referred by two
different misleading names. Reading and comparing the file content we
see that the properly named RFC index has these characteristics:

\begin{enumerate}

    \item an embedded date at line 4, \texttt{7/14/1994},
    
    \item the \textbf{right sentence},

    \item the list of citations \textbf{reverse sorted, from the newest
        to the oldest}.
    
\end{enumerate}

Instead, the misnamed RFC indexes have:

\begin{enumerate}

    \item the \textbf{wrong sentence},
    
    \item the list of citations \textbf{sorted from the oldest to the
        newest}.

\end{enumerate}

The above findings suggests that the \texttt{rfc-index.txt} file was
probably created on 14 July 1994, as the embedded date suggests, and
copied on the FTP server the day after; it could be useful if the list
of citations showed further evidence of the above, but this can not
happen, since the file begins with RFC0001 and then it is corrupted well
before it reaches RFCs issued by July 1994. 

We assume that the bug was then introduced and the RFC files generated
afterwards show the unbalanced parenthesis.

To support this deduction, we see that the wrong files have a date on
the FTP server that is 15 July 1994, as it is reported in table
\vref{table:indexesmetadata}; even assuming that the date was changed
for some reason the last RFC shown on \texttt{std-index.txt} --- the
only file that is not corrupted --- is RFC1653, that was issued on July
1994.  So, in any case, the \texttt{std-index.txt} file could not have
been generated before the begin of July 1994.  To summarize:

\begin{enumerate}

    \item it is highly probable that a correct RFC index was generated
        on 14 July 1994\footnote{We have further evidence that on
        December 1993 the RFC index was correct, so there is support
        that before 1994 the mistake was not present.},
    
    \item a mistaken file, that had been necessarily created on July
        1994 or later, is present on the FTP server of the Clausthal
        University of Technology with a timestamp of 15 July 1994.

\end{enumerate}

In our opinion the most plausible scenario is that the
\texttt{rfc-index.txt} files generated before 14 July 1994 had no error,
that the mistake was introduced that day and thereafter the answer to
the question posed at the beginning of this paper is: \textit{an
unbalanced parenthesis may go unnoticed for more that twenty-six years}.


\section{Discussion} 
\label{sec:Discussion}

There is an empirical law --- dubbed Linus'~law --- that states that
``given enough eyeballs, all bugs are shallow''.  It was formulated by
Eric S.  Raymond in the book ``The Cathedral and the Bazaar'' and so
named in honour of Linus Torvalds, Linux creator.

The law applies to software, not to documentation, and it has been
criticized so there is no clear evidence either of its validity or
its falsity. 

Our little investigation would like to bring some evidence towards a
better understanding of the idea behind the Linus'~law: is it true that
simply having a content under a wide public scrutiny ensures for its
quality ? If we compare syntax errors on documentation to software bugs
in computer code --- which is a not too-far stretched analogy, in our
opinion --- then the present paper gives a negative answer.

We have tried to understand if there is some \textit{other} reason
behind the fact that the error went unfixed for so long, and among the
issues we noticed that:

\begin{itemize}

    \item it is not easy to provide a proper feedback for this kind of
        error: it is possible to provide a RFC
        errata\footnote{\url{https://www.rfc-editor.org/errata.php}},
        but it is limited to RFCs and the last resort is emailing the
        \url{mailto:rfc-editor@rfc-editor.org} address --- we did that
        on 10 of January 2021;

    \item it is not explained how the indexes are created: we have not
        been able to find the repository of the software that generates
        them and file a bug report.

\end{itemize}

In our opinion the main enabler is not the ``number of eyeballs'', to
quote the law statement, but \textit{how easy it is to contribute
changes}. Clearly open source and free software have this property, the
same does not necessarily hold for documentation, even if it is freely
available (legally speaking, the RFCs licenses are very permissive) and
freely distributable.


\section{Resources} 
\label{sec:Resources}

\subsection{Links} 
\label{sub:Links}

\begin{description}

    \item[TU Clausthal's FTP server] The anonymous FTP server is
        reachable at the address:
        \url{ftp://ftp.tu-clausthal.de/pub/docs/rfc/}

    \item[rfc-index.txt]
        \url{ftp://ftp.tu-clausthal.de/pub/docs/rfc/other_indexes/rfc-index.txt}

    \item[fyi-index.txt]
        \url{ftp://ftp.tu-clausthal.de/pub/docs/rfc/other_indexes/fyi-index.txt}

    \item[std-index.txt]
        \url{ftp://ftp.tu-clausthal.de/pub/docs/rfc/standards/std-index.txt}

\end{description}


\subsection{Indexes} 
\label{sub:Indexes}

Excerpt of the header of the index files analyzed in this paper.

\begin{description}

    \item[rfc-index.txt]\noindent
{
\scriptsize
\begin{verbatim}
                           # RFC INDEX #
                           -------------

                             7/14/1994

This file contains citations for all RFCs in reverse numeric order.  RFC
citations appear in this format:

NUM STD    Author 1, ... Author 5., "Title of RFC",  Issue date.
         (Pages=##) (Format=.txt or .ps)  (FYI ##) (STD ##) (RTR ##)
            (Obsoletes RFC####) (Updates RFC####)

Key to citations:

    #### is the RFC number; ## p. is the total number of pages.

    The format and byte information follows the page information in
    parenthesis.  The format, either ASCII text (TXT) or PostScript (PS) or
    both, is noted, followed by an equals sign and the number of bytes for
    that version (Post- Script is a registered trademark of Adobe Systems
    Incorporated).  The example (Format: PS=xxx TXT=zzz bytes) shows that
    the PostScript version of the RFC is xxx bytes and the ASCII text version
    is zzz bytes.

    The (Also FYI ##) phrase gives the equivalent FYI number if the RFC was
    also issued as an FYI document.

    "Obsoletes xxx" refers to other RFCs that this one replaces; "Obsoleted
    by xxx" refers to RFCs that have replaced this one.  "Updates xxx" refers
    to other RFCs that this one merely updates (but does not replace);
    "Updated by xxx" refers to RFCs that have been updated by this one (but
    not replaced).  Only immediately succeeding and/or preceding RFCs are
    indicated, not the entire history of each related earlier or later RFC
    in a related series.
\end{verbatim}
}

    \item[fyi-index.txt]\noindent
{
\scriptsize
\begin{verbatim}
~~~~~~~~~~~~~~~~~~~~~~~~~~~~~~~~~~~~~~~~~~~~~~~~~~~~~~~~~~~~~~~~~~~~

                             RFC INDEX
                           -------------

This file contains citations for all RFCs in numeric order.

RFC citations appear in this format:

####  Title of RFC.  Author 1, Author 2, Author 3.  Issue date.
      (Format: ASCII) (Obsoletes xxx) (Obsoleted by xxx) (Updates xxx)
      (Updated by xxx) (Also FYI ####)

Key to citations:

#### is the RFC number.

Following the number are the title (terminated with a period), the
author, or list of authors (terminated with a period), and the date
(terminated with a period).

The format and byte information follows in parenthesis.  The format,
either ASCII text (TXT) or PostScript (PS) or both, is noted, followed
by an equals sign and the number of bytes for that version.  For
example (Format: TXT=aaaaa, PS=bbbbbb bytes) shows that the ASCII text
version is aaaaa bytes, and the PostScript version of the RFC is
bbbbbb bytes.

Obsoletes xxxx refers to other RFCs that this one replaces;
Obsoleted by xxxx refers to RFCs that have replaced this one.
Updates xxxx refers to other RFCs that this one merely updates but
does not replace); Updated by xxxx refers to RFCs that have been
updated by this one (but not replaced).  Only immediately succeeding
and/or preceding RFCs are indicated, not the entire history of each
related earlier or later RFC in a related series.
\end{verbatim}
}

    \item[std-index.txt]\noindent
{
\scriptsize
\begin{verbatim}
~~~~~~~~~~~~~~~~~~~~~~~~~~~~~~~~~~~~~~~~~~~~~~~~~~~~~~~~~~~~~~~~~~~~

                             RFC INDEX
                           -------------

This file contains citations for all RFCs in numeric order.

RFC citations appear in this format:

####  Title of RFC.  Author 1, Author 2, Author 3.  Issue date.
      (Format: ASCII) (Obsoletes xxx) (Obsoleted by xxx) (Updates xxx)
      (Updated by xxx) (Also FYI ####)

Key to citations:

#### is the RFC number.

Following the number are the title (terminated with a period), the
author, or list of authors (terminated with a period), and the date
(terminated with a period).

The format and byte information follows in parenthesis.  The format,
either ASCII text (TXT) or PostScript (PS) or both, is noted, followed
by an equals sign and the number of bytes for that version.  For
example (Format: TXT=aaaaa, PS=bbbbbb bytes) shows that the ASCII text
version is aaaaa bytes, and the PostScript version of the RFC is
bbbbbb bytes.

Obsoletes xxxx refers to other RFCs that this one replaces;
Obsoleted by xxxx refers to RFCs that have replaced this one.
Updates xxxx refers to other RFCs that this one merely updates but
does not replace); Updated by xxxx refers to RFCs that have been
updated by this one (but not replaced).  Only immediately succeeding
and/or preceding RFCs are indicated, not the entire history of each
related earlier or later RFC in a related series.
\end{verbatim}
}

\end{description}



\section{Acknowledgments} 
\label{sec:Acknowledgments}

The author thanks Andrea \textit{`ap'} Paolini and Guido \textit{`zen'}
Bolognesi for their detective skills and TU Clausthal's system
administrators for keeping up a piece of the old Internet.


\end{document}